\documentclass[aps,manuscript,showpacs]{revtex4}
\usepackage{graphicx}

\begin{document}
\widetext

\title{Anti-Hyperon polarization in high energy $pp$ collisions with polarized beams}
\author{Ye Chen$^a$, Zuo-tang Liang$^a$, Ernst Sichtermann$^b$, Qing-hua Xu$^{a,b}$ and Shan-shan Zhou$^a$} 
\affiliation{$^a$Department of Physics, Shandong University, Jinan, Shandong 250100, China\\
$^b$Nuclear Science Division, MS 70R0319,
Lawrence Berkeley National Laboratory, Berkeley, CA 94720}

\date{\today}

\begin{abstract}

We study the longitudinal polarization of the $\bar \Sigma^-$, $\bar \Sigma^+$, $\bar \Xi^0$ and $\bar \Xi^+$ anti-hyperons  
in polarized high energy $pp$ collisions at large transverse momenta, extending a recent study for the $\bar \Lambda$ anti-hyperon.
We make predictions by using different parameterizations of the polarized parton densities and 
models for the polarized fragmentation functions.
Similar to the $\bar \Lambda$ polarization, the $\bar \Xi^0$ and $\bar \Xi^+$ polarizations are found to be sensitive to 
the polarized anti-strange sea,  $\Delta \bar s(x)$,  in the nucleon.
The $\bar \Sigma^-$ and $\bar \Sigma^+$ polarizations show sensitivity to the light sea quark polarizations, 
$\Delta \bar u(x)$ and $\Delta \bar d(x)$, and their asymmetry.

\end{abstract}


\maketitle


\section{Introduction}

The self spin-analyzing parity violating decay~\cite{tdlee} of  hyperons and anti-hyperons provides a practical way to determine 
the hyperon and anti-hyperon polarization by measuring the angular distributions of the decay products.
The polarizations have been used widely in studying various aspects of spin physics in high energy 
reactions~\cite{aleph,opal,nomad,hermes,e665,compass}.
The discovery of transverse hyperon polarization in unpolarized hadron-hadron and hadron-nucleus collisions in the 1970s 
led to many subsequent studies, both experimentally and theoretically~\cite{transExp,transTh}.
Phenomenological studies of longitudinal hyperon polarization may be found in 
Refs.~\cite{Burkardt:1993zh,GH93,BL98,Kot98,Florian98,MS98,LL00,LXL01,
XLL02,LL02,XL04,DZL05,Boros:2000ex,MSY,Anselmino:2000ga,Ellis2002,xls}.
The main themes of these studies can be categorized as follows.
On the one hand, one aims to study spin transfer in the fragmentation process.
On the other, one aims to get insight in the spin structure of the initial hadrons.
Experimental data are available from $e^+e^-$-annihilation at the $Z$-pole~\cite{aleph,opal}, 
deep-inelastic scattering with polarized beams and targets~\cite{nomad,hermes,e665,compass}, 
and proton-proton $(pp)$ collisions~\cite{xu05}.

The study of anti-hyperon polarization remains topical.  
Recent COMPASS data~\cite{compass} seem to indicate a difference between Lambda and anti-Lambda polarization.  
The proton spin physics program at the Relativistic Heavy Ion Collider (RHIC) has come online~\cite{RHIC}.  
A recent study for anti-Lambda polarization in polarized pp collisions~\cite{xls} shows sensitivity to the anti-strange quark spin contribution to the proton spin, which is only poorly constrained by existing data. 
In this paper, we evaluate the longitudinal polarization of 
the $\bar \Sigma^-$, $\bar \Sigma^+$, $\bar \Xi^0$ and $\bar \Xi^+$ anti-hyperons in $pp$ collisions 
at large transverse momenta $p_T$.
Section II contains the phenomenological framework and discusses the current knowledge of the parton distribution and fragmentation functions.  
The contributions to the anti-hyperon production cross sections are discussed in Section III, 
followed by results for the polarizations in Section IV and a short summary in Section V.

\section{Phenomenological Framework}

The method to calculate the longitudinal polarization of high $p_T$ hyperons and anti-hyperons in 
high energy $pp$ collisions is based on the factorization theorem and perturbative QCD and has been described in earlier works~\cite{BL98,LL00,LXL01,XLL02,LL02,XL04,DZL05}.
For self-containment we briefly summarize the key elements and emphasize the aspects that are specific 
to anti-hyperons in the following of this section. 

\subsection{Formalism} 
We consider the inclusive production of high $p_T$ anti-hyperons ($\bar{H}$) in $pp$ collisions with one of 
the beams longitudinally polarized.
The longitudinal polarization of $\bar H$ is defined as,
\begin{equation}
P_{\bar{H}}(\eta)\equiv \frac
{d\sigma{(p_+p \to  \bar{H}_+X)}-d\sigma{(p_+p \to  \bar{H}_-X)}}
{d\sigma{(p_+p \to  \bar{H}_+X)}+d\sigma{(p_+p \to  \bar{H}_-X)}}
= \frac {d\Delta \sigma}{d\eta} (\vec pp \to  \bar{H} X) /
\frac {d\sigma}{d\eta}(pp \to  \bar{H} X),
\label{gener1}
\end{equation}
where $\eta$ is the pseudo-rapidity of the $\bar{H}$, the subscripts $+$ and $-$ denote 
positive and negative helicity, 
and $\Delta\sigma$ and $\sigma$ are the polarized and unpolarized inclusive production cross sections.

We assume that transverse momentum $p_T$ of the produced $\bar H$ is high enough so that the cross section can be factorized in a collinear way. 
In this case, the $\bar{H}$'s come merely from the fragmentations of high $p_T$ partons from $2\to 2$ 
hard scattering ($ab \to cd$) with one initial parton polarized, and 
the polarized inclusive production cross section is given by,
\begin{equation}
\frac {d\Delta \sigma}{d\eta}{(\vec pp \to  \bar{H} X)}
=\int_{p_T^{min}}dp_T
\sum_{abcd}\int dx_a dx_b \Delta f_a(x_a)f_b(x_b)
D_L^{\vec ab\to \vec cd}(y)\frac {d \hat {\sigma}}{d\hat t}{( ab\to cd)}
\Delta D_c^{\bar{H}}(z),
\label{dsig}
\end{equation}
where the transverse momentum $p_T$ of $\bar H$ is integrated above a threshold $p_T^{min}$;
the sum concerns all sub-processes; 
$\Delta f_a(x_a)$ and $f_b(x_b)$ are the polarized and unpolarized parton distribution functions in the proton,
$x_a$ and $x_b$ are the momentum fractions carried by partons $a$ and $b$,
$D_{L}^{\vec ab\to\vec cd}(y)\equiv d\Delta\hat\sigma/d\hat\sigma$ is 
the partonic spin transfer factor in the elementary hard process $\vec ab\to\vec cd$ with cross section $\hat{\sigma}$,
$y\equiv p_b \cdot (p_a-p_c)/p_a \cdot p_b$ is defined in terms of the four momenta $p$ of the partons $a$--$d$, 
and  $\Delta D_c^{\bar{H}}(z)$ is the polarized fragmentation function.
It is defined by,
\begin{equation}
\Delta D_c^{\bar{H}}(z) \equiv D_{c}^{\bar{H}}(z,+)-D_{c}^{\bar{H}}(z,-),
\label{pff}
\end{equation}
in which the argument $z$ is the momentum carried by $\bar{H}$ relative to the momentum of the fragmenting parton $c$, 
and the arguments $+$ and $-$ denote that the produced $\bar{H}$ has equal or opposite helicity as parton $c$.
The scale dependencies of the parton distribution and fragmentation functions have been omitted for notational clarity.
Intrinsic transverse momenta in the proton and in the fragmentation process are 
small compared to $p_T$ and are not considered above.

The unpolarized inclusive production cross section, $d\sigma/d\eta$, is given by an analogous expression with unpolarized 
parton distribution and fragmentation functions.

\subsection{Inputs}
The differential cross section is a convolution of three factors, the parton distribution functions, 
the cross section of the elementary hard process, and the fragmentation functions. 
We discuss them separately in the following.

\subsubsection{The partonic spin transfer factors $D_L^{\vec ab\to \vec cd}(y)$ in the hard scattering}
The partonic spin transfer factor $D_L^{\vec ab\to\vec cd}(y)$ is
determined by the spin dependent hard scattering cross sections, many of which are known at leading and next to leading order~\cite{gastmans,bms,jssv}.
The leading order results for $D_L^{\vec ab\to\vec cd}(y)$ are functions only of $y$, defined above, and are tabulated in Ref.~\cite{XLL02} for quarks.
Here, we are particularly interested in anti-quarks.
By using charge conjugation, a symmetry
which is strictly valid in QCD, one obtains
\begin{equation}
D_L^{\vec{\bar q}_1q_2\to \vec{\bar q}_1q_2}(y)=D_L^{\vec{q}_1q_2\to \vec{q}_1q_2}(y),
\end{equation}
\begin{equation}
D_L^{\vec{\bar q}_1g\to\vec{\bar q}_1g}(y)=D_L^{\vec{q}_1g\to \vec{q}_1g}(y),
\end{equation}
etc.
Consistency demands that the partonic spin transfer be used with the polarized parton distributions and fragmentation functions of the same order.
In view of the current knowledge of the polarized fragmentation functions, we consider only the leading order.

\subsubsection{The parton distribution functions}
\begin{figure}[htb!]
\includegraphics[width=12cm]{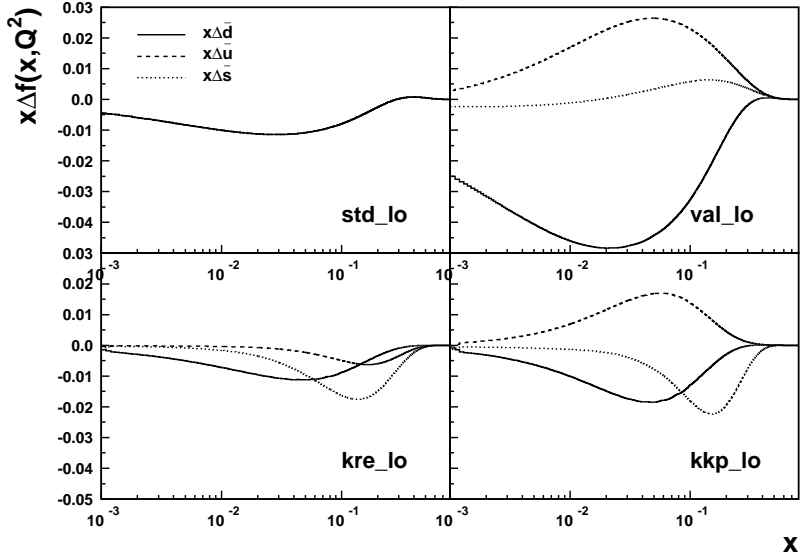}
\caption{Polarized anti-sea quark distributions from the leading order GRSV2000 (upper) and DNS2005 (lower) 
parameterizations evaluated at a scale $Q=10\,\mathrm{GeV}$.}
\label{grsv2000}
\end{figure}

The unpolarized parton distribution functions $f(x)$ are determined from unpolarized deep inelastic scattering 
and other related data from unpolarized experiments.  
Many parametrizations exist also for the polarized parton distribution functions $\Delta f(x)$, 
e.g. GRSV2000, BB, LSS, GS, ACC, DS2000, and DNS2005~\cite{pol_grsv,pol_bb,pol_lss,pol_gs,pol_acc,pol_ds,pol_dns}.
However, they are much less well constrained by data and large differences exist in particular for 
the parameterizations of the polarized anti-sea distributions.
This is illustrated in Fig.~\ref{grsv2000}, where the anti-sea distributions from the GRSV2000~\cite{pol_grsv} and DNS2005~\cite{pol_dns} parametrizations 
are shown.

\subsubsection{The polarized fragmentation functions $\Delta D_c^{\bar{H}}(z)$}\label{sec:model}
The polarized fragmentation functions $\Delta D_c^{\bar{H}}(z)$ are defined in Eq.~(\ref{pff}) and represent the difference of the probability densities for finding an anti-hyperon $\bar{H}$ in a parton $c$ with equal and opposite helicity as parton $c$.
They have been studied experimentally in $e^+e^-$-annihilation, polarized deep-inelastic scattering, and high $p_T$ hadron production in polarized $pp$ collisions, but present data\cite{aleph,opal,nomad,hermes,e665,compass,xu05} do not provide satisfactory constraints.
It is thus necessary to make an \emph{Ansatz} or model.

Several \emph{Ans\"atze} have been adopted in the literature~\cite{Kot98,Florian98,MS98,MSY}.
One possibility is to relate $\Delta D_c^{\bar{H}}(z)$ to the unpolarized fragmentation functions $D_c^{\bar{H}}(z)$ by means of  spin transfer coefficients.
Alternatively, one could use the Gribov relation~\cite{Gribov}, a proportionality relating $D_c^{\bar{H}}(z) \propto q_c^{\bar{H}}(z)$ and $\Delta D_c^{\bar{H}}(z) \propto \Delta q_c^{\bar{H}}(z)$ in which $q_c^{\bar{H}}(z)$ and $\Delta q_c^{\bar{H}}(z)$ are the parton distributions in the (anti-)hyperon.
Other approaches to $\Delta D_c^{\bar{H}}(z)$ are in essence often different combinations or approximations of the aforementioned approaches.
Although this paper is focused on the leading order, it should be noted that in general the distributions depend also on the scale $Q^2$.
The $Q^2$-evolutions of the distributions are generally different and \emph{Ans\"atze} of this type are thus necessarily made at specific values of $Q^2$.
The $\bar{H}$ sample produced in experiment consists of the QCD contributions and of contributions from electromagnetic or electroweak decays of heavier resonances.
The latter are not formally part of the cross sections in the factorized framework, but do need to be considered in comparisons between theory and experiment.

In this paper, we use the Lund string fragmentation model~\cite{pythia} to describe the hadronization process and hence to evaluate the probability densities expressed by the fragmentation functions.
This model enables us to distinguish between directly produced $\bar{H}$ and $\bar{H}$ which are decay products of heavier resonances and, furthermore, to distinguish between $\bar{H}$ that contain the initial parton $c$ and those that do not.
It should be noted that Lund string fragmentation is in general not equivalent to the independent fragmentation in the factorized framework, and the fragmentation probability densities are not necessarily universal.
We expect that such effects are numerically unimportant for the $\bar{H}$ considered in this paper, $\bar{H}$ produced at central rapidities with large $p_T$ in polarized $pp$ collisions at RHIC, and have verified explicitly for the $\bar{\Lambda}$ in these conditions that indeed the same (anti-)quark fragmentation probabilities are found as in a kinematically equivalent colliding $e^+e^-$ system.

Keeping the notation, we thus express polarized fragmentation function as the sum of contributions from directly produced and from decay $\bar{H}$,
\begin{equation}
\Delta D_c^{\bar{H}}(z)=\Delta D_c^{\bar{H}}(z;\mathrm{direct})+\Delta D_c^{\bar{H}}(z;\mathrm{decay}),
\end{equation}
and evaluate the decay contribution $\Delta D_c^{\bar{H}}(z;\mathrm{decay})$ by a sum over parent anti-hyperons $\bar{H}_j$ and kinematic convolutions,
\begin{equation}
\Delta D_c^{\bar{H}}(z;\mathrm{decay})=\sum_j\int dz' t^D_{\bar{H},\bar{H}_j}K_{\bar{H},\bar{H}_j}(z,z')\Delta D_c^{\bar{H}_j}(z'),
\label{eq:dec}
\end{equation}
in which the kernel function $K_{\bar{H},\bar{H}_j}(z,z')$ is the probability for the decay of the parent $\bar{H}_j$ 
with a fractional momentum $z'$ to
produce $\bar{H}$ with fractional momentum $z$, and $t^D_{\bar{H},\bar{H}_j}$ is the spin transfer factor in the decay process.

Most of the decay processes involved here are two body decay $\bar H_j\to \bar H \bar M$. 
For an unpolarized two body decay, the kernel function $K_{\bar H,\bar H_j}(z,z')$ can easily be calculated 
and is the same as that for $H_j\to H M$.
In this case, the magnitude of the momentum of $H$ is fixed and the direction is isotropically distributed in the rest frame 
of the parent $H_j$. A Lorentz transformation to the moving frame of $H_j$ gives,
\begin{equation}
K_{H,H_j}(z,\vec p_T;z',\vec p_T\ ')=\frac{N}{E_j}Br(H_j\to H M)\delta(p\cdot p_j-m_jE^*),
\label{decayK}
\end{equation}
where $Br(H_j\to HM)$ is the decay branching ratio, $N$ is a normalization constant, and $E^*$ is the energy of $H$ 
in the rest frame of $H_j$, which depends on the parent mass $m_j$ and on the masses of the decay products.
The corresponding calculation of $K_{H,H_j}(z,z')$ for a {\it polarized} $H_j$ is more involved since the angular decay  
distribution can be anisotropic in the case of a weak decay and each decay process needs to be dealt with separately.
However, since the $E^*$ is usually small compared to the momentum of $H_j$ in the $pp$ center of
mass frame, the anisotropy can be neglected and Eq.(\ref{decayK}) forms a good approximation.

The decay spin transfer factors for hyperons $H_j\to H+X$ are discussed and given e.g. in Refs.(\cite{GH93,LXL01}).
Charge conjugation is a good symmetry for these decays, and hence $t^D_{\bar{H},\bar{H}_j}=t^D_{H,H_j}$.
Furthermore, $\Delta D_c^{\bar{H}}(z;{\rm direct})=\Delta D_{\bar c}^H(z;{\rm direct})$, so that only $\Delta D_c^{H}(z;\mathrm{direct})$ need to be modeled.
The main strong decay contributions are those from $J^P=(3/2)^+$ hyperons, such as $\Sigma^*\to\Sigma\pi$, and $\Xi^*\to\Xi\pi$.
The electromagnetic and weak decay contributions, for example $\Sigma^0\to\Lambda\gamma$ and $\Xi\to\Lambda\pi$, can be evaluated analogously and we will include these contributions in our results.

Our aforementioned classification follows Refs.~\cite{GH93,BL98,LL00,LXL01,XLL02,LL02} and distinguishes (A) directly produced hyperons that contain the initial quark $q$ of flavor $f$; (B) decay products of heavier polarized hyperons; (C) directly produced  hyperons that do not contain the initial $q$; (D) decay products of heavier unpolarized hyperons.
Therefore, 
\begin{equation}
D_f^{H}(z;\mathrm{direct})=D_f^{H(\mathrm{A})}(z)+D_f^{H(\mathrm{C})}(z).
\end{equation}
for the unpolarized fragmentation functions and, similarly, for the polarized fragmentation functions,
\begin{equation}
\Delta D_f^{H}(z;\mathrm{direct})=\Delta D_f^{H(\mathrm{A})}(z)+\Delta D_f^{H(\mathrm{C})}(z).
\end{equation}
We assume that directly produced hyperons which do not contain the initial quark are unpolarized, so that
\begin{equation}
\Delta D_f^{H(\mathrm{C})}(z)=0.
\label{eq:DfC}
\end{equation}
The polarization of directly produced hyperons then originates only from category (A) and is given by,
\begin{equation}
\Delta D_f^{H(\mathrm{A})}(z)=t^F_{H,f} D_f^{H(\mathrm{A})}(z),
\label{eq:DfA}
\end{equation}
in which $t^F_{H,f}$ is known as the fragmentation spin transfer factor.
If the quarks and anti-quarks produced in the fragmentation process are unpolarized, consistent with Eq.(\ref{eq:DfC}), 
then $t^F_{H,f}$ is a constant given by,
\begin{equation}
t_{H,f}^F=\Delta Q_f/n_f,
\label{eq:tHf}
\end{equation}
where $\Delta Q_f$ is the fractional spin contribution of a quark with flavor $f$ to the spin of the hyperon,  
and $n_f$ is the number of valence quarks of flavor $f$ in $H$.
In recursive cascade hadronization models, such as Feynman-Field type fragmentation models \cite{ff78}  where a simple elementary process takes place recursively, $D_f^{H(A)}(z)$ and $D_f^{H(C)}(z)$ are well defined and determined.
In such hadronization models, $D_f^{H(A)}(z)$ is the probability to produce a first rank $H$ with fractional momentum $z$.
This probability is usually denoted by $f_{q_f}^H(z)$ in these models, so that $D_f^{H(C)}(z)=D_f^H(z;\mathrm{direct})-f_{q_f}^H(z)$, and $f_{q_f}^H(z)$ is well determined by unpolarized fragmentation data.
Hence, in such models the $z$-dependence of the polarized fragmentation functions $\Delta D$ is obtained from the unpolarized fragmentation functions, which are empirically known.
The only unknown is the spin transfer constant $t^F_{H,q}=\Delta Q_f/n_f$. 
By using either the SU(6) wave function or polarized deep-inelastic lepton-nucleon scattering data, two distinct expectations have been made for $\Delta Q_f$, the so-called SU(6) and DIS expectations~\cite{LL00}.

The approach described above has been applied to the polarizations of different hyperons in $e^+e^-$, semi-inclusive DIS and $pp$ collisions, anti-hyperons in semi-inclusive DIS and anti-Lambda in $pp$ \cite{GH93,BL98,LL00,LXL01,XLL02,LL02,XL04,DZL05,xls}.
The results can be compared with data~\cite{aleph,opal,nomad,hermes,e665,compass,xu05}.
The current experimental accuracy does not allow one to distinguish between the expectations for $t^F_{H,f}$ based on the SU(6) and DIS pictures.
The $z$-dependence of the available data on $\Lambda$ polarization is well described~\cite{LL00}.
The approach is thus justified by existing data.
We will use both the SU(6) and DIS pictures for our present predictions.

\subsection{Implementation} 
The expression for the polarization of anti-hyperons, $P_{\bar H}$, follows from the definition in Eq.(\ref{gener1}) and the factorized cross sections (c.f. Eq.(\ref{dsig})), and is given by
\begin{equation}
P_{\bar{H}}(\eta)=\frac
{\int_{p_T^{min}}dp_T\sum_{abcd}\int dx_a dx_b f_a(x_a)f_b(x_b)
\frac {d \hat {\sigma}}{d\hat t}{( ab\to cd)}P_{c/ab\to cd}(x_a,y)\Delta D_c^{\bar{H}}(z)}
{\int_{p_T^{min}}dp_T\sum_{abcd}\int dx_a dx_b f_a(x_a)f_b(x_b)\frac{d\hat{\sigma}}{d\hat t}{( ab\to cd)} D_c^{\bar{H}}(z)},
\label{pol}
\end{equation}
where, $P_{c/ab\to cd}(x_a,y)=D_L^{\vec ab\to\vec cd}(y)\Delta f_a(x_a)/f_a(x_a)$ is the polarization of parton $c$ before it fragments.
This can be rewritten using the model to calculate $\Delta D_c^{\bar H}(z)$ according to the origin of $\bar H$ (c.f. section~\ref{sec:model}),
\begin{equation}
P_{\bar{H}}(\eta)=\frac
{\int_{p_T^{min}}dp_T\sum_{abcd}\sum_{\alpha}\int dx_a dx_b f_a(x_a)f_b(x_b)
\frac {d \hat {\sigma}}{d\hat t}{( ab\to cd)}D_c^{\bar{H}(\alpha)}(z)P_{c/ab\to cd}(x_a,y)S_c^{\bar H(\alpha)}(z)}
{\int_{p_T^{min}}dp_T\sum_{abcd}\int dx_a dx_b f_a(x_a)f_b(x_b)\frac{d\hat{\sigma}}{d\hat t}{( ab\to cd)} D_c^{\bar{H}}(z)},
\label{pol2}
\end{equation}
where the summation over $\alpha$ concerns the four process classes from which the $\bar H$ originate, and $S_c^{\bar H(\alpha)}(z)=\Delta D_c^{\bar H(\alpha)}(z)/D_c^{\bar H(\alpha)}(z)$ denotes the spin transfer factor in the fragmentation for each of these classes (c.f. section~\ref{sec:model}).

The anti-hyperon polarization $P_{\bar {H}}$ can be in principle be evaluated numerically from Eq.~(\ref{pol}).
In practice, it is feasible and actually advantageous to use a Monte-Carlo event generator that incorporates all hard scattering processes and uses a recursive hadronization model to evaluate $P_{\bar {H}}$ from Eq.~(\ref{pol2}).
In particular, this ensures in a natural way that all sub-processes, including feed-down contributions, are taken into account.
Furthermore, it facilitates comparisons of the predictions with experiment data since event generator output can be propagated through detailed detector simulations so that, for example, the experiment kinematic acceptance for the observed $\bar H$ decay products can be taken into account without relying on extrapolation over unmeasured regions.

We have used the {\sc pythia} event generator, which incorporates the hard scattering processes and uses the Lund string fragmentation model~\cite{pythia}, in our calculations.
{\sc pythia} is commonly used for hadron-hadron collisions and its output has been tested and tuned to describe a vast body of data. 
In particular, reasonable agreement is found for $\Lambda+\bar {\Lambda}$ production that was recently measured for transverse momenta up to 5\,GeV in $pp$ collisions at RHIC~\cite{STAR2007} when a $K$-factor of $\sim3$ is used.
For the results presented here we have used {\sc pythia} version 6.4 with all hard scattering processes selected and initial and final state radiation switched off.
We have verified that the aforementioned $K$-factor does not affect the polarization results.

\section {Anti-hyperon production}

The longitudinal polarization of high $p_T$ anti-hyperons produced in $pp$ collisions is determined by the polarization of 
the initial partons taking part in the hard scattering, the partonic spin transfer factor, and the spin transfer in 
the fragmentation process.
Since the up, down, and strange quark and anti-quark polarizations in the polarized proton are different and 
the spin transfer in the fragmentation process for a given type of anti-hyperon is flavor dependent as well, 
the contributions to $\bar{H}$ production from the fragmentation of different quark flavors and gluons need to be studied.
These contributions are independent of polarization and have been determined in multi-particle production data in high energy reactions.
An impressive body of data has been collected over the past decades and the contributions can thus be 
considered to be known accurately and to be well-modeled in Monte-Carlo event generators.

As described in Section II C, we have used the {\sc pythia} generator~\cite{pythia} to evaluate the contributions to the production of the $\bar \Sigma^-$, 
$\bar \Sigma^+$, $\bar \Xi^0$ and $\bar \Xi^+$ anti-hyperons.
The flavor compositions of these anti-hyperons lead us to expect a large contribution to the production of 
$\bar \Sigma^+(\bar d\bar d\bar s)$ from $\bar d$-fragmentation, a large contribution to $\bar \Sigma^-(\bar u\bar u\bar s)$ 
production from $\bar u$-fragmentation, and a large contribution to $\bar \Xi$-production from $\bar s$-fragmentation.

\begin{figure}[h!]
\includegraphics[width=16cm,height=10cm]{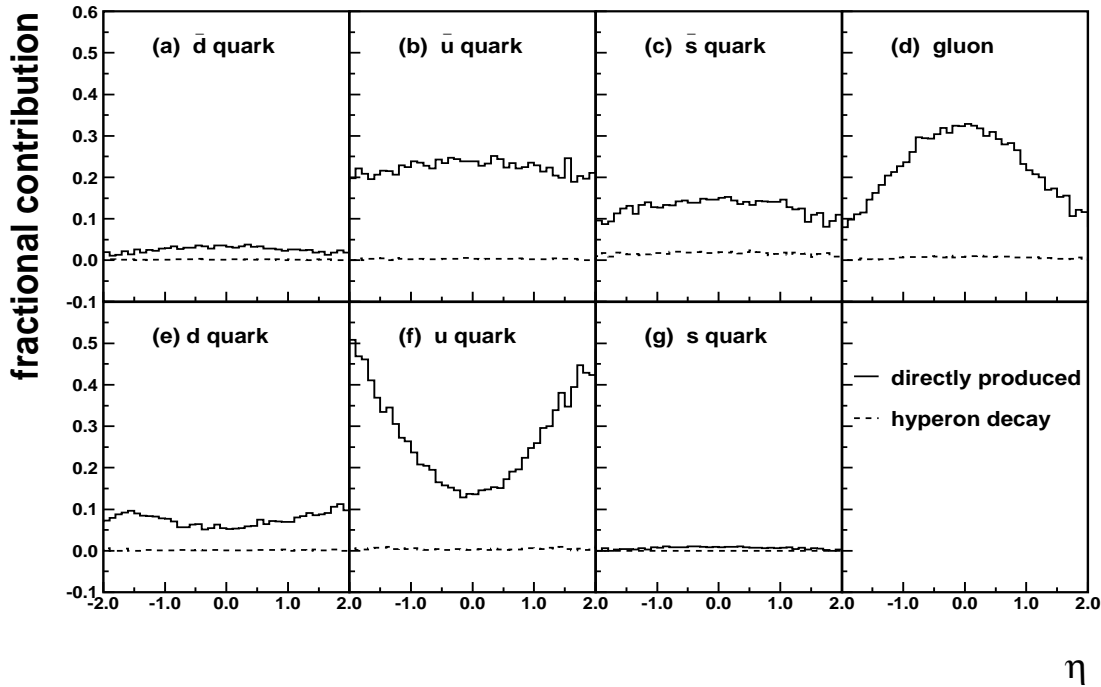}
\caption{Contributions to $\bar\Sigma^-(\bar u \bar u \bar s)$ production with $p_T\ge 8$ GeV/c
in $pp$ collisions at $\sqrt s=200$ GeV.
The continuous and dashed lines are respectively
the directly produced and decay contributions.}
\label{sigmabar-_eta}
\end{figure}

\begin{figure}[h!]
\includegraphics[width=16cm,height=10cm]{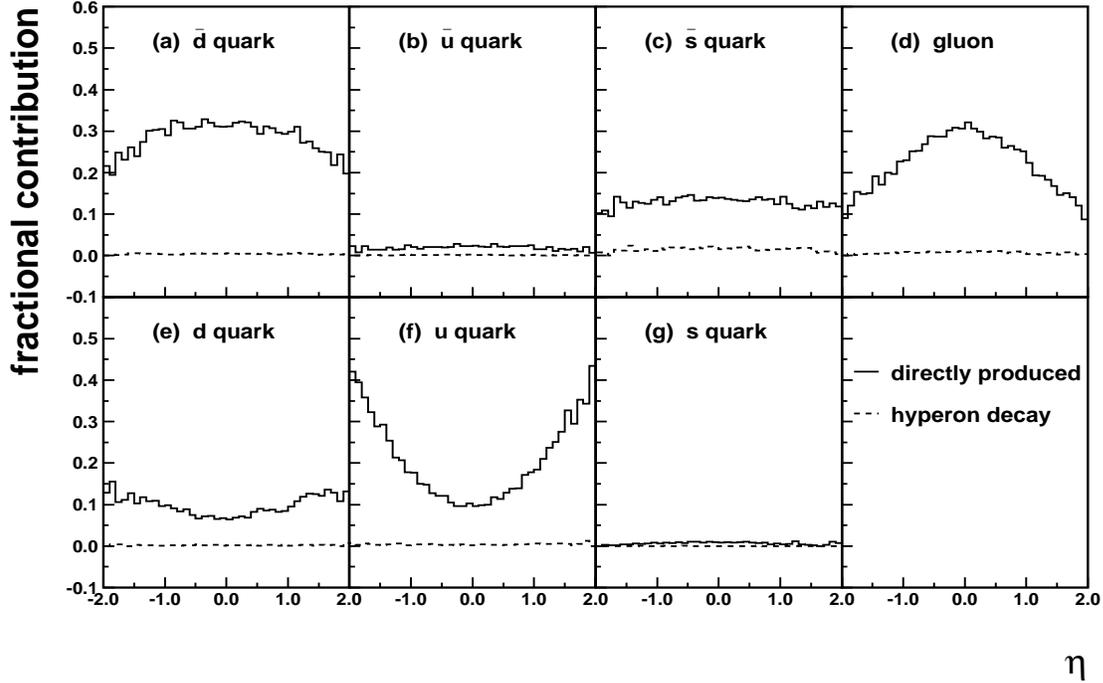}
\caption{Contributions to $\bar\Sigma^+(\bar d \bar d \bar s)$ production with $p_T\ge 8$ GeV/c
in $pp$ collisions at $\sqrt s=200$ GeV.
The continuous and dashed lines are respectively
the directly produced and decay contributions.}
\label{sigmabar+_eta}
\end{figure}

Fig.~\ref{sigmabar-_eta} shows the results for the fractional contributions to $\bar\Sigma^-$ production from 
the fragmentation of anti-quarks/quarks of different flavors and of gluons in $pp$ collisions 
at $\sqrt{s} = 200\,\mathrm{GeV}$ for hyperon transverse momenta $p_T > 8\,\mathrm{GeV}$ versus pseudo-rapidity $\eta$.
The decay contribution to $\bar\Sigma^-$ production is seen to be negligibly small.
This is different than for $\bar\Lambda$ production and implies that its polarization measurement will reflect 
more directly the spin structure of the nucleon and the polarized fragmentation function.
The results are symmetric in $\eta \leftrightarrow -\eta$ since $pp$ collisions are considered.
The $\bar d$-quark and $s$-quark fragmentation contributions originate from second or higher rank particles 
in the fragmentation and have similar shapes since both are sea quarks (anti-quarks).
The increasingly large $u$-quark contribution with increasing $|\eta|$ originates from valence quarks.
Only the $\bar u$-quark and $\bar s$-quark give first rank fragmentation contributions.
These contributions are sizable.
Most important for the production of $\bar\Sigma^-$ with $p_T > 8\,\mathrm{GeV}$ and $|\eta| < 1$ are $\bar u$ 
and gluon fragmentation.
Since the $\bar\Sigma^-$ spin is carried mostly by the $\bar u$-quark spins, this implies that the
$\bar\Sigma^-$ polarization in singly polarized $pp$ collisions should be sensitive to $\Delta \bar u(x)$, 
the $\bar u$-quark polarization distribution in the polarized proton.

The results for $\bar\Sigma^+$, shown in Fig.~\ref{sigmabar+_eta}, are similar to those for $\bar\Sigma^-$ 
when $\bar u$ and $\bar d$ are interchanged.
The small difference between the fractional contribution of the $\bar u$-quark to $\bar \Sigma^-$ production 
and of the $\bar d$-quark to $\bar \Sigma^+$ production reflects the asymmetry of the light sea density in the proton, 
$\bar d(x)$$>$$\bar u(x)$, which is built into the parton distribution functions.
The large $\bar d$-quark fragmentation contribution to $\bar\Sigma^+$ production and the large $\bar d$-quark 
spin contribution to the $\bar\Sigma^+$ spin lead us to expect that $\bar\Sigma^+$ polarization measurements in $pp$ 
collisions are sensitive to $\Delta \bar d(x)$.

\begin{figure}[h!]
\includegraphics[width=16cm,height=10cm]{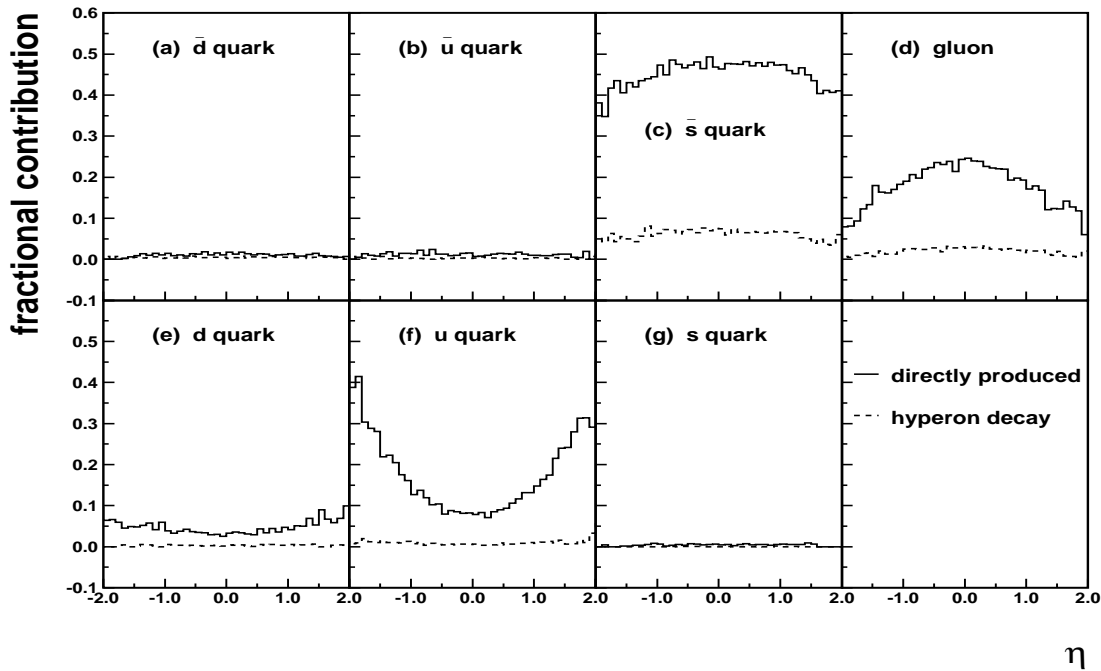}
\caption{Contributions to $\bar \Xi^0(\bar u \bar s \bar s)$ 
production with $p_T\ge 8$ GeV/c in $pp$ collisions at $\sqrt s=200$ GeV.
The continuous and dashed lines are respectively
the directly produced and decay contributions.}
\label{xibar0_eta}
\end{figure}

\begin{figure}[h!]
\includegraphics[width=16cm,height=9.8cm]{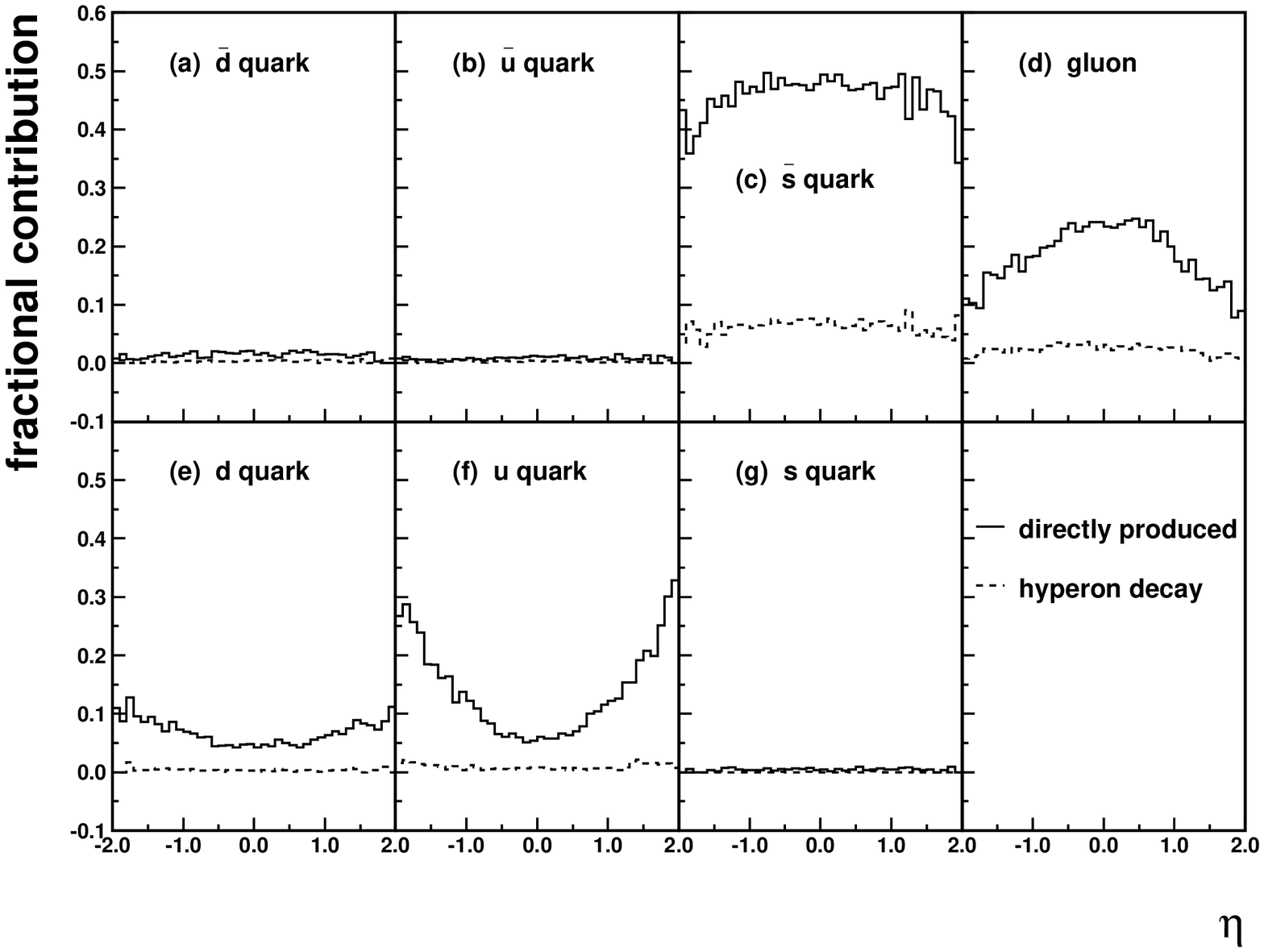}
\caption{Contributions to $\bar \Xi^+(\bar d \bar s \bar s)$ 
production with $p_T\ge 8$ GeV/c in $pp$ collisions at $\sqrt s=200$ GeV.
The continuous and dashed lines are respectively
the directly produced and decay contributions.}
\label{xibarp_eta}
\end{figure}

\begin{figure}[h!]
\includegraphics[width=15cm,height=8cm]{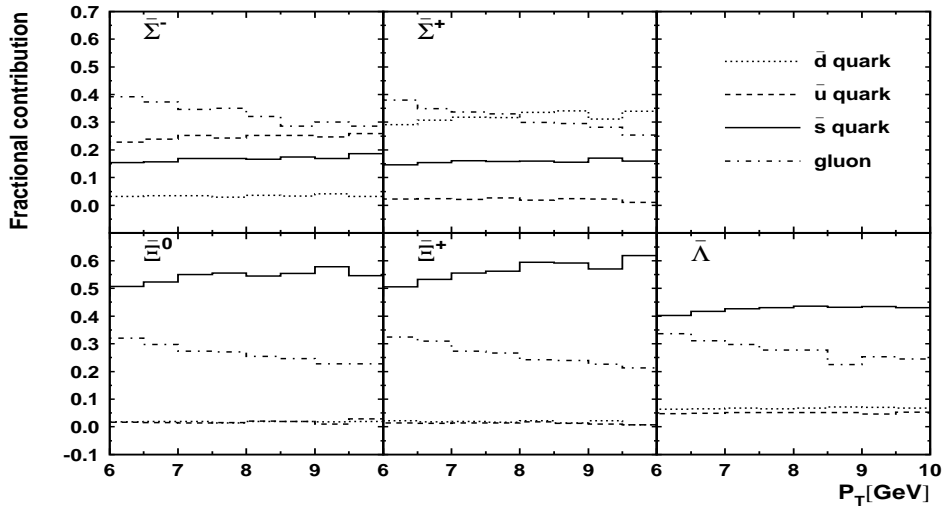}
\caption{Contributions to $\bar\Sigma^-(\bar u\bar u \bar s)$, $\bar \Sigma^+(\bar d \bar d \bar s)$,
$\bar \Xi^0(\bar u \bar s \bar s)$ and $\bar \Xi^+(\bar d \bar s \bar s)$ production for 
$|\eta|<1$ in $pp$ collisions at $\sqrt s=200$ GeV versus transverse momentum $p_T$.}
\label{hyorg_pt}
\end{figure}

Figs.~\ref{xibar0_eta} and \ref{xibarp_eta} show the fractional contributions for $\bar \Xi^0$ and $\bar \Xi^+$ production, 
respectively.
The dominant contribution originates from $\bar s$-quark fragmentation. 
It amounts to almost half the $\bar \Xi$ production, and is larger than the $\bar d$-quark fragmentation contribution 
to $\bar\Sigma^+$ production and the  $\bar u$-quark fragmentation contribution to $\bar\Sigma^-$ production.
This results from strange suppression, which reduces the relative contributions from $\bar u$ and $\bar d$-quark fragmentation 
to $\bar\Xi$ production.
We thus expect that  $\bar\Xi$ polarization measurements are sensitive to $\Delta \bar s(x)$ in the nucleon.
They are thus complementary to polarization measurements of the $\bar\Lambda$~\cite{xls}, which has a larger production 
cross section but also larger decay contributions.

In Fig.~\ref{hyorg_pt} we show the $p_T$ dependence of the fractional fragmentation contributions for the mid-rapidity region, 
$|\eta| < 1$, for the $\bar\Sigma^-$, $\bar\Sigma^+$, $\bar\Xi^0$, $\bar\Xi^+$, as well as the $\bar\Lambda$.
The anti-quark contributions generally increase with increasing $p_T$, and the gluon contributions decrease.

\section {Results and discussion}

\begin{figure}
\includegraphics[width=16cm,height=10cm]{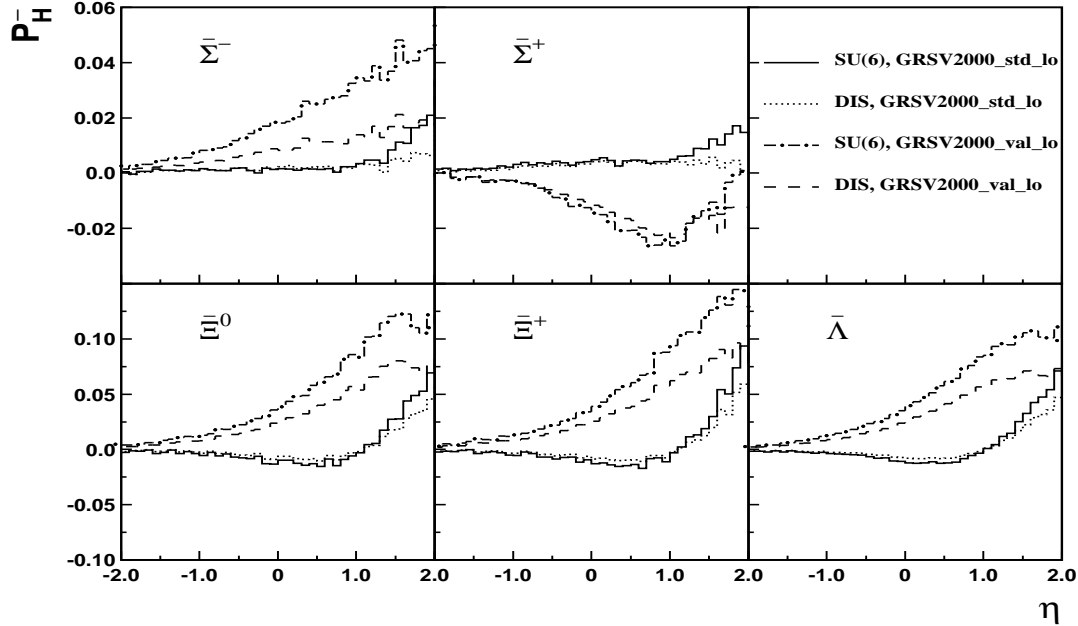}
\caption{Longitudinal polarization for anti-hyperons 
with transverse momentum $p_T\ge 8$ GeV/c in $pp$ collisions at $\sqrt s=200$ GeV 
with one longitudinally polarized beam versus pseudo-rapidity $\eta$. 
Positive $\eta$ is taken along the direction of the polarized beam.}
\label{hpol_eta}
\end{figure}

We have evaluated $P_{\bar{H}}$ for the $\bar{\Sigma}^-$, $\bar{\Sigma}^+$, $\bar{\Xi}^0$, and $\bar{\Xi}^+$ anti-hyperons 
as a function of $\eta$ for $p_T\ge 8$GeV and $\sqrt{s} = 200\,\mathrm{GeV}$ using different parametrizations for 
the polarized parton distributions and using the SU(6) and DIS pictures for the spin transfer factors $t^F_{H,q}$ in the fragmentation.
In all cases, the unpolarized parton distributions of Ref.~\cite{GRV98} were used.
The results using the polarized parton distributions of Ref.~\cite{pol_grsv} are shown in Fig.~\ref{hpol_eta}, 
together with our previous results for $P_{\bar{\Lambda}}$~\cite{xls}.
The main characteristics are:
\begin{itemize}
\item{the size of the polarization increases in the forward direction with respect to the polarized proton beam and can be as large 
as ~10\% ($\bar{\Xi}^0$, $\bar{\Xi}^+$) at $\eta = 2$,}
\item{the differences between the $\bar{H}$ polarizations obtained for different parametrizations of the polarized parton 
distribution functions are generally larger than the differences between the results for different models for 
the spin transfer in fragmentation,}
\item{the size of the polarizations for the $\bar \Sigma^-$ and $\bar \Sigma^+$ hyperons is smaller than for 
the $\bar\Lambda$ and $\bar\Xi$ hyperons because of the lower fractional contributions from $\bar{u}$ 
and $\bar{d}$ fragmentation to $\bar \Sigma^-$ and $\bar \Sigma^+$ production than from $\bar{s}$ fragmentation to 
the $\bar\Lambda$ and $\bar\Xi$ production,}
\item{the results for $\bar \Sigma^-$ and $\bar \Sigma^+$ for the GRSV2000 valence distributions differ in sign 
because of the sign difference in $\Delta \bar u(x)$ and $\Delta\bar d(x)$, and in size and shape because of 
flavor-symmetry breaking in the unpolarized and this polarized parton distribution scenario,}
\item{the $\bar \Xi^0$ and $\bar \Xi^+$ polarizations are similar to each other because of the dominance 
of $\bar{s}$-fragmentation; they are somewhat larger than the $\bar\Lambda$ polarization because of the smaller 
decay contributions and their sensitivity to $\Delta\bar s$ is thus more direct.}
\end{itemize}

\begin{figure}
\includegraphics[width=16cm,height=10cm]{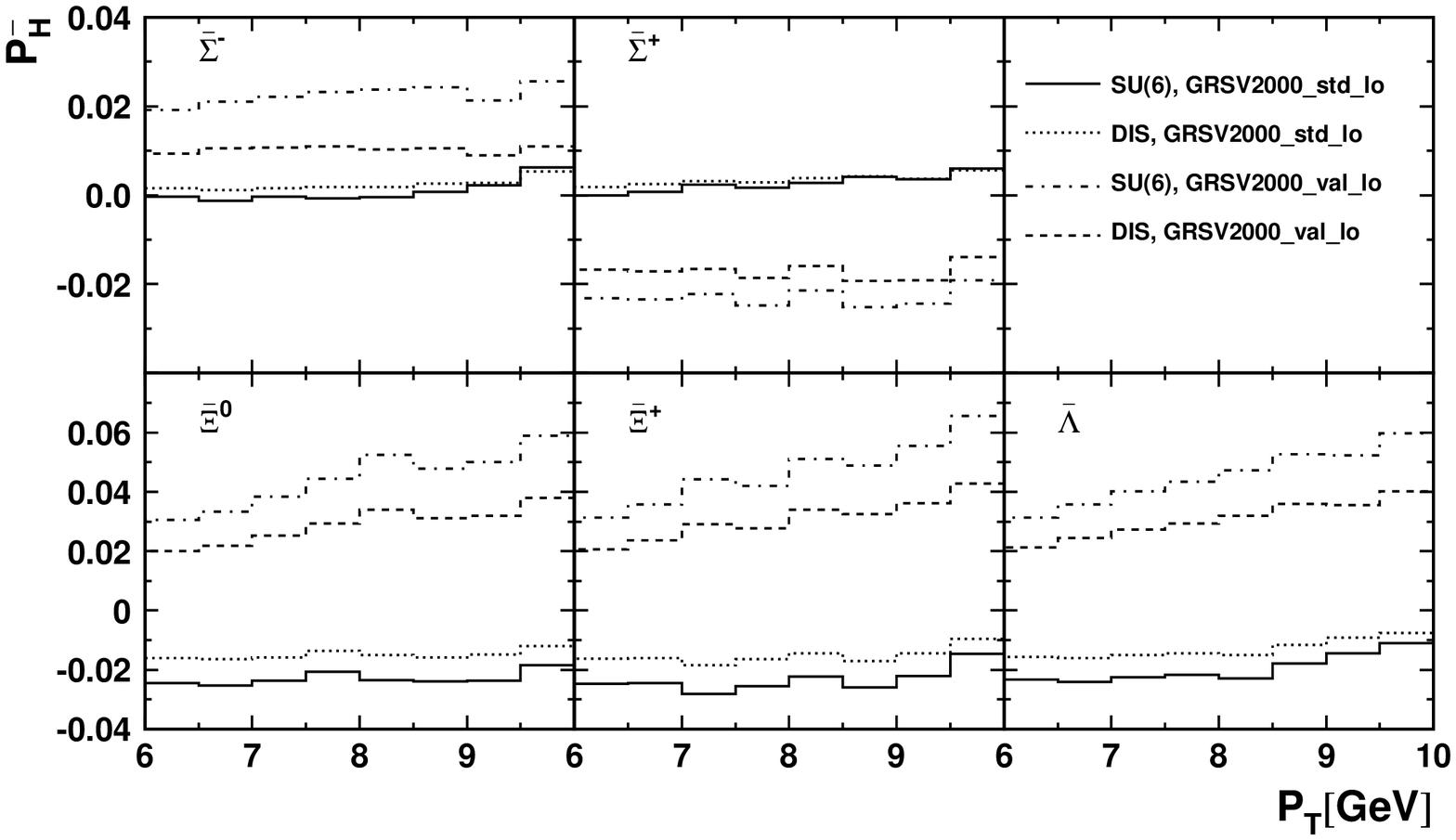}
\caption{Longitudinal polarization for anti-hyperons 
with pseudo-rapidity $0 < \eta <1$ in $pp$ collisions at $\sqrt s=200$ GeV
with one longitudinally polarized beam versus transverse momentum $p_T$.
Positive $\eta$ is taken along the direction of the polarized beam.}
\label{hpol_pt}
\end{figure}

\begin{figure}
\includegraphics[width=16cm,height=10cm]{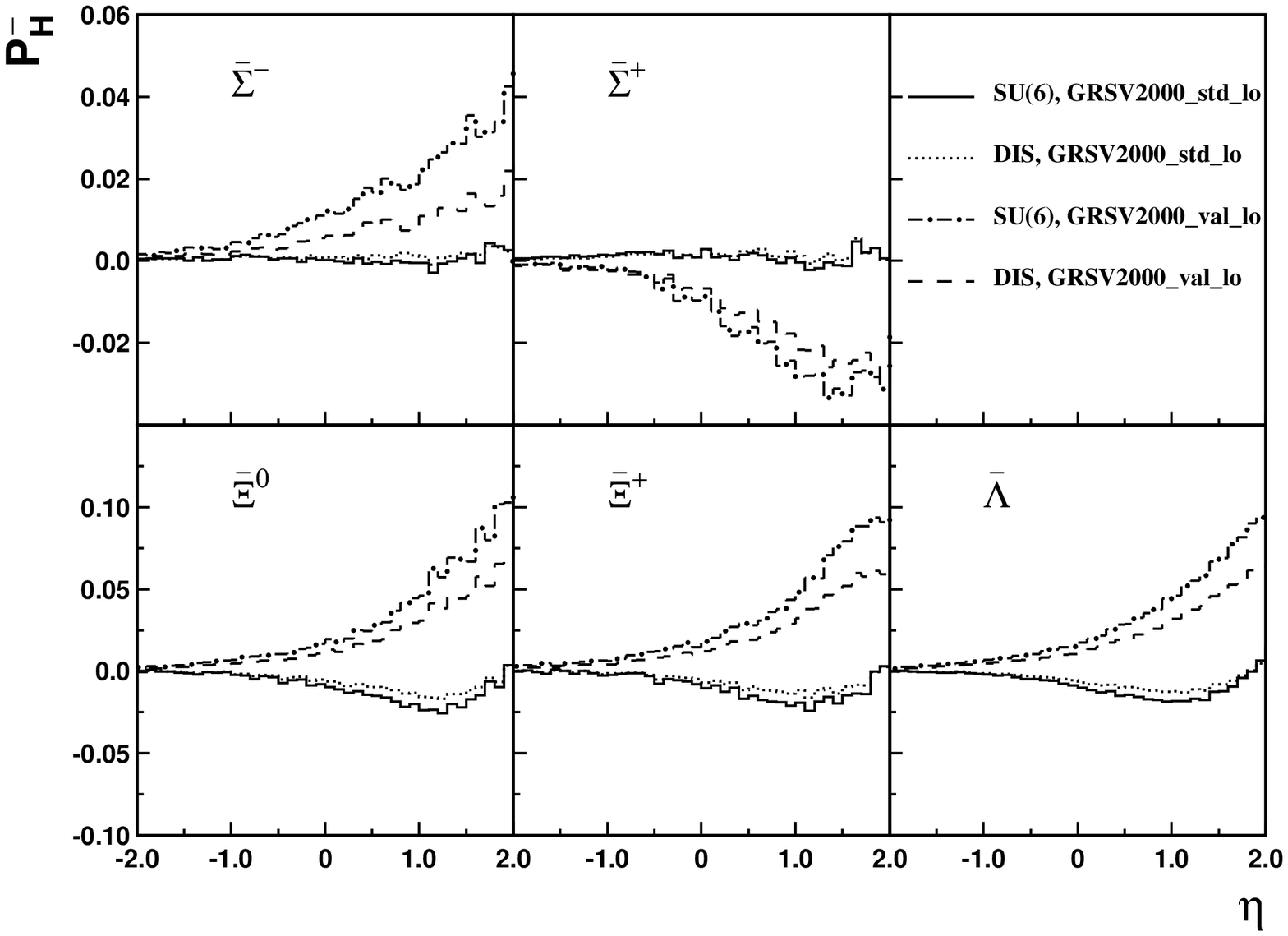}
\caption{Longitudinal polarization for anti-hyperons with $p_T>10$ GeV 
as a function of the pseudo-rapidity 
$\eta$ in $pp$ collisions at $\sqrt s=500$ GeV
with one longitudinally polarized beam. 
Positive $\eta$ is taken along the direction of the polarized beam.}
\label{hpol500}
\end{figure}

Fig.~\ref{hpol_pt} shows the polarizations in the pseudo-rapidity range $0 < \eta<1$ versus transverse momentum $p_T$.
The polarizations are sensitive mostly to the polarized anti-quark distributions for momentum fractions $0.05 < x < 0.25$ 
and the $p_T$-dependences are consequently not very strong.
Only a modest variation is expected also with center-of-mass energy.
To illustrate this, we have repeated the calculations for $\sqrt{s}=500$ GeV.
Fig.~\ref{hpol500} shows the $\eta$-dependence for $p_T > 10\,\mathrm{GeV}$.
Apart from differences expected from phase-space, the results are seen to be very similar to those in Fig.~\ref{hpol_eta}.

Gluon fragmentation is seen to contribute sizably to the production of anti-hyperons in the central rapidity range in Figs.(2-6).
The fragmentation of gluons is less well known than that of quarks even in the unpolarized case and also the polarization of gluons in the colliding polarized protons cannot be determined precisely from current data.
In the estimates, as in earlier calculations~\cite{GH93,BL98,LL00,LXL01,XLL02,LL02,XL04,DZL05,xls},  we have taken into account the spin transfer in the hard scattering of gluons, and have neglected the spin transfer in the fragmentation of gluons into anti-hyperons.
This assumption is consistent with the model used for the polarized fragmentation functions.
Data with improved precision on the production cross sections and on gluon polarization would provide important constraints.
In particular, we expect that our results are largely unaffected if the gluon spin contribution to the proton spin is found to be small.
Current data~\cite{gluonExp} excludes the large values that were proposed originally to explain the quark spin contribution to the proton spin~\cite{anomaly}.

Last, we have estimated the precision with which $\bar{H}$ polarization measurements could be made at RHIC~\cite{RHIC}.
For an analyzed integrated luminosity of  $\mathcal{L} \simeq 300\,\mathrm{pb}^{-1}$ and a proton beam polarization 
of $P \simeq 70\%$, we anticipate that e.g. $P_{\bar{\Xi}}$ could be measured to within $\sim$0.02 uncertainty.
Measurements of $\bar{H}$ polarization at RHIC are thus worthwhile in view of the presently limited knowledge 
of $\Delta\bar q(x)$ in the nucleon, as evidenced in particular by the large differences between 
the parametrization sets of Ref.~\cite{pol_grsv}.

\section{summary}
In summary, we have evaluated the longitudinal polarizations of the $\bar\Sigma^-$, $\bar\Sigma^+$, $\bar\Xi^0$, and $\bar\Xi^+$ 
anti-hyperons in highly energetic collisions of longitudinally polarized proton beams.
The results show sensitivity to the anti-quark polarizations in the nucleon sea.
In particular, the $\bar\Sigma^-$ and $\bar\Sigma^+$ polarizations are sensitive to the light sea quark polarizations, 
$\Delta\bar u(x)$ and $\Delta\bar d(x)$.
The $\bar\Xi^0$ and $\bar\Xi^+$ polarizations are sensitive to strange anti-quark polarization $\Delta\bar s(x)$.
Precision measurements at the RHIC polarized $pp$-collider should be able to provide new insights in the sea 
quark polarizations in the nucleon.

\vspace{0.3cm}
{\bf Acknowledgments}
This work was supported in part by the National Science Foundation of China (NSFC) under Grant Nos. 10525523 and 10405016, 
and by the United States Department of Energy under Contract No. DE-AC02-05CH11231, Office of Nuclear Physics, 
and by the Department of Science and Technology of Shandong Province of China under Contract
No. 2006GG2210013.
.


\begin{thebibliography}{99}
\bibitem{tdlee}T. D. Lee and C. N. Yang, Phys. Rev. {\bf 108}, 1645 (1957); 
               T.D. Lee, J. Steinberger, G.Feinberg, P.K. Kabir, and C.N. Yang, Phys. Rev. {\bf 106}, 1367 (1957).
\bibitem{aleph} D. Buskulic {\it et al.} [ALEPH Collaboration],
              Phys. Lett. B {\bf 374}, 319 (1996).
\bibitem{opal} K. Ackerstaff {\it et al.} [OPAL Collaboration],
               Eur. Phys. J. C {\bf 2}, 49 (1998).
\bibitem{nomad} P. Astier {\it et al}. [NOMAD Collaboration],
              Nucl. Phys. B {\bf 588}, 3 (2000); {\bf 605}, 3 (2001).
\bibitem{hermes}A. Airapetian {\it et al.} [HERMES Collaboration],
          Phys.Rev. D {\bf 64}, 112005 (2001); {\bf 74}, 072004 (2006).
\bibitem{e665} M. R. Adams {\it et al.} [E665 Collaboration], Eur. Phys. J. C{\bf 17}, 263 (2000).
\bibitem{compass} M. G. Sapozhnikov, [COMPASS Collaboration],
  hep-ex/0503009 and hep-ex/0602002; also V.Yu. Alexakhin, [COMPASS Collaboration], hep-ex/0502014.\
\bibitem{transExp} A. Lesnik {\it et al.,} Phys. Rev. Lett. {\bf 35}, 770 (1975);
G.~Bunce {\it et al.}, Phys. Rev. Lett. {\bf 36}, 1113 (1976);
K. Heller {\it et al.}, Phys. Rev. Lett. {\bf 41}, 607 (1978);
For a review, see e.g., A.~Bravar, in Proc. of the 13th International Symposium
on High Energy Spin Physics, Protvino, Russia, September 1998, 
edited by N.E.~Tyurin {\it et al},  World Scientific, Singapore, 1999, p.167.
\bibitem{transTh} See e.g., B.~Andersson, G.~Gustafson and G.~Ingelman, Phys. Lett. {\bf 85B}, 417 (1979);  
     T.A.~DeGrand and H.I.~Miettinen, Phys. Rev. D {\bf 24}, 2419 (1981); 
     J.~Soffer and N.~T\"ornqvist, Phys. Rev. Lett. {\bf 68}, 907 (1992); 
     Z. T. Liang and C. Boros, Phys. Rev. Lett. {\bf 79}, 3608 (1997); Phys. Rev. D {\bf 61}, 117503 (2000); 
     H. Dong and Z. T. Liang, Phys. Rev. D{\bf 70}, 014019 (2004), and references therein. 
\bibitem{Burkardt:1993zh}M.~Burkardt and R.~L.~Jaffe, Phys.\ Rev.\ Lett.\  {\bf 70}, 2537 (1993); \
  R. L. Jaffe, Phys. Rev. D {\bf 54}, 6581 (1996).
\bibitem{GH93} G. Gustafson and J. H\"{a}kkinen, Phys. Lett. B {\bf 303}, 350 (1993).
\bibitem{BL98} C. Boros and Z. T. Liang, Phys. Rev. D {\bf 57}, 4491 (1998).
\bibitem{Kot98} A. Kotzinian, A. Bravar and D. von Harrach, Eur. Phys. J. C{\bf 2}, 329 (1998).
\bibitem{Florian98} D. de Florian, M. Stratmann, and W. Vogelsang,
             Phys. Rev. Lett. {\bf 81}, 530 (1998); \
             Phys. Rev. D{\bf 57}, 5811 (1998).
\bibitem{MS98} B.~Q.~Ma and J.~Soffer,  Phys.\ Rev.\ Lett.\  {\bf 82}, 2250 (1999).
\bibitem{LL00} C. X. Liu and Z. T. Liang, Phys. Rev. D {\bf 62}, 094001 (2000).
\bibitem{LXL01} C. X. Liu, Q. H. Xu and Z. T. Liang, Phys. Rev. D {\bf 64}, 073004 (2001).
\bibitem{XLL02} Q. H. Xu, C. X. Liu and Z. T. Liang, Phys. Rev. D {\bf 65}, 114008 (2002).
\bibitem{LL02} Z. T. Liang and C. X. Liu, Phys. Rev. D {\bf 66}, 057302 (2002).
\bibitem{XL04} Q. H. Xu and Z. T. Liang,  Phys. Rev. D {\bf 70}, 034015 (2004).
\bibitem{DZL05}H.~Dong, J.~Zhou and Z.~T.~Liang, Phys.\ Rev.\ D {\bf 72}, 033006 (2005).
\bibitem{Boros:2000ex}C.~Boros, J.~T.~Londergan and A.~W.~Thomas, Phys.\ Rev.\  D {\bf 62}, 014021 (2000).
\bibitem{MSY}B.~Q.~Ma, I.~Schmidt and J.~J.~Yang, Phys.\ Rev.\ D {\bf 61}, 034017 (2000); {\bf 63}, 037501 (2001);
B.~Q.~Ma, I.~Schmidt, J.~Soffer and J.~J.~Yang, Phys.\ Rev.\ D {\bf 62}, 114009 (2000);
 Eur.\ Phys.\ J.\ C {\bf 16}, 657 (2000); Nucl.\ Phys.\ A {\bf 703}, 346 (2002).
\bibitem{Anselmino:2000ga}M.~Anselmino, M.~Boglione and F.~Murgia,
  Phys.\ Lett.\  B {\bf 481}, 253 (2000).
\bibitem{Ellis2002} J.~R.~Ellis, A.~Kotzinian and D.~V.~Naumov,
     Eur.\ Phys.\ J.\ C {\bf 25}, 603 (2002); J.~Ellis, A.~Kotzinian, D.~Naumov and M.~Sapozhnikov, 
     Eur. Phys. J. C {\bf 52},283 (2007).
\bibitem{xls}Q. H. Xu, Z. T. Liang, and E. Sichtermann,
 Phys. Rev. D {\bf 73}, 077503 (2006).
\bibitem{xu05} Q. H. Xu [STAR collaboration], hep-ex/0512058 and hep-ex/0612035.
\bibitem{RHIC}
  G.~Bunce, N.~Saito, J.~Soffer and W.~Vogelsang,
  Ann.\ Rev.\ Nucl.\ Part.\ Sci.\  {\bf 50}, 525 (2000).
\bibitem{gastmans}R. Gastmans and T.T. Wu, ÒThe Ubiquitous PhotonÓ (Clarendon Press - Oxford, 1990).
\bibitem{bms}J. Babcock, E. Monsay, D. W. Sivers, Phys. Rev. Lett. {\bf 40}, 1161 (1978); \ Phys. Rev. D {\bf19}, 1483 (1979).
\bibitem{jssv}B. J\"{a}ger, A. Sch\"{a}fer, M. Stratmann, W. Vogelsang, Phys. Rev. D {\bf 67}, 054005 (2003).
\bibitem{pol_grsv}M. Gl\"uck, E. Reya, M. Stratmann
and W. Vogelsang, Phys. Rev. D {\bf 63}, 094005 (2001); \  Phys. Rev. D {\bf 53}, 4775 (1996).
\bibitem{pol_bb}J. Bl\"{u}mlein and H. B\"{o}ttcher, Nucl. Phys. B {\bf 636}, 225 (2002).
\bibitem{pol_lss}E. Leader, A. V. Sidorov and D. B. Stramenov,
Phys. Rev. {\bf D} 73, 034023 (2006); {\it ibid} {\bf D} 75, 074027 (2007).
\bibitem{pol_gs}T. Gehrmann and W. J. Stirling, Phys. Rev. D {\bf 53}, 6100 (1996).
\bibitem{pol_acc}M. Hirai {\it et al.} [Asymmetry Analysis Collaboration] 
 Phys. Rev. D {\bf 69}, 054021 (2004); {\it ibid} D {\bf 74}, 014015 (2006).
\bibitem{pol_ds}D. De Florian and R. Sassot, Phys. Rev. D {\bf 62}, 094025 (2000).
\bibitem{pol_dns}D. De Florian, G. A. Navarro, R. Sassot, Phys. Rev. D {\bf 71}, 094018 (2005).
\bibitem{Gribov} V. N. Gribov, and L. N. Lipatov, Phys. Lett. B {\bf 37}, 78 (1971); \
                 Yad.\ Fiz.\  {\bf 15}, 1218 (1972)
                 [Sov.\ J.\ Nucl.\ Phys.\  {\bf 15}, 675 (1972)].
\bibitem{pythia} T. Sj\"{o}strand, S. Mrenna, P. Skands, JHEP {\bf 0605}, 026 (2006);
  B. Andersson, G. Gustafson, G. Ingelman and T. Sj\"{o}strand, Phys. Rep. {\bf 97}, 31 (1983).
\bibitem{ff78}   R.~D.~Field and R.~P.~Feynman,
  Nucl.\ Phys.\  B {\bf 136}, 1 (1978).
\bibitem{STAR2007}   B.~I.~Abelev {\it et al.}  [STAR Collaboration],
  Phys.\ Rev.\  C {\bf 75}, 064901 (2007).
\bibitem{GRV98} M. Gl\"uck, E. Reya, and A. Vogt, Eur. Phys. J. C {\bf 5}, 461 (1998).
\bibitem{gluonExp} S. Adler {\it et al.}(PHENIX Collaboration), 
Phys. Rev. Lett. {\bf 93}, 202002 (2004),  Phys. Rev. D{\bf 73}, 091102 (2006), 
A. Adare {\it et al.}(PHENIX Collaboration), Phys. Rev. D{\bf 76}, 051106 (2007);
B.I. Abelev {\it et al.} (STAR Collaboration), 
Phys. Rev. Lett. {\bf 97}, 252001 (2006), and arXiv:0710.2048 [hep-ex](2007);
E.~S.~Ageev {\it et al.}  [COMPASS Collaboration], Phys.\ Lett.\  B {\bf 633}, 25 (2006), {\bf 647}, 8 (2007); 
B.~Adeva {\it et al.}  [Spin Muon Collaboration (SMC)]  Phys.\ Rev.\  D {\bf 70}, 012002 (2004);
A.~Airapetian {\it et al.}  [HERMES Collaboration], Phys.\ Rev.\ Lett.\  {\bf 84}, 2584 (2000).
\bibitem{anomaly} G. Altarelli and G.G. Ross, Phys.\ Lett.\ B {\bf 212}, 391 (1988); R.D. Carlitz, J.C. Collins, and A.H.~Mueller, Phys.\ Lett.\ B {\bf 214}, 229 (1988).

\end{thebibliography}
\end{document}